\documentclass[12pt,preprint]{aastex}
\usepackage{psfig,natbib}
\shortauthors{Bower et al.}
\shorttitle{Linear Polarization of Sgr~A*}
\begin{document}

\newcommand\degd{\ifmmode^{\circ}\!\!\!.\,\else$^{\circ}\!\!\!.\,$\fi}
\newcommand{\etal}{{\it et al.\ }}
\newcommand{\uv}{(u,v)}
\newcommand{\rdm}{{\rm\ rad\ m^{-2}}}
\newcommand{\msuny}{{\rm\ M_{\sun}\ y^{-1}}}
\newcommand{\mylesssim}{\stackrel{\scriptstyle <}{\scriptstyle \sim}}


\title{Interferometric Detection of Linear Polarization from Sagittarius A* at 230 GHz}

\author{Geoffrey C. Bower\altaffilmark{1}, 
Melvyn C.H. Wright\altaffilmark{1},
Heino Falcke\altaffilmark{2}, 
Donald C. Backer\altaffilmark{1} }

\altaffiltext{1}{Astronomy Department \& Radio Astronomy Laboratory, 
University of California, Berkeley, CA 94720; gbower,wright,dbacker@astro.berkeley.edu}
\altaffiltext{2}{Max Planck Institut f\"{u}r Radioastronomie, Auf dem 
H\"{u}gel 69, D 53121 Bonn Germany; hfalcke@mpifr-bonn.mpg.de} 

\begin{abstract}

We measured the linear polarization of Sagittarius A* to be $7.2 \pm 0.6 \%$ at 230 GHz
using the BIMA array with a resolution of $3.6 \times 0.9$ arcsec.  
This confirms the previously reported detection with the JCMT 14-m antenna.
Our high resolution observations demonstrate that the polarization does 
not arise from dust but from a synchrotron 
source associated with Sgr A*.  We see no change in the polarization
position angle and only a small
change in the polarization fraction in four observations distributed over 60 days.
We find a position angle $139 \pm 4$ degrees that differs substantially
from what was found in earlier JCMT observations at the same frequency.  
Polarized dust emission cannot account for this discrepancy leaving variability and
observational error as the only explanations.
The BIMA observations alone place an upper limit on the magnitude of the rotation measure of 
$2\times 10^6 \rdm$.  These new observations when combined with the JCMT observations 
at 150, 375 and 400 GHz suggest 
${\rm RM}=-4.3 \pm 0.1 \times 10^5 \rdm$.  This RM may be caused by an external
Faraday screen.
Barring a special geometry or a high number of field reversals, 
this RM rules out accretion rates greater than $\sim 10^{-7} \msuny$.  This measurement is
inconsistent with high accretion rates necessary in standard advection dominated
accretion flow and Bondi-Hoyle models for Sgr A*.  It argues for low accretion rates
as a major factor in the overall faintness of Sgr A*.

\end{abstract}

\keywords{Galaxy: center --- galaxies: active --- polarization --- radiation
mechanisms: non-thermal --- scattering }

\section{Introduction}

Linear and circular polarization are important diagnostics of synchrotron sources.  In
active galactic nuclei and galactic microquasars,
polarization is useful for characterizing particle energy, magnetic
field configuration and jet composition 
\citep{1985ApJ...298..114M,1998Natur.395..457W,
2002MNRAS.336...39F}.
Variability in polarization is important for
diagnosing shocks and the stability of accretion processes.

Here we present new observations of linear and circular polarization in Sagittarius A*,
the compact nonthermal radio source in the Galactic Center.  Sgr A* is associated with a
$3 \times 10^6 {\rm M_{\sun}}$ black hole \citep{2001ARA&A..39..309M}.  
While much is known about this source, there is still dispute over the nature of
accretion and outflow in the source.  Advection dominated accretion flows (ADAFs),
convection dominated accretion flows (CDAFs), Bondi-Hoyle accretion, Blandford-K\"onigl-type jets
and hybrids of these ideas have been used to characterize the spectrum and size of Sgr A*
\citep[e.g.,][]{1998ApJ...492..554N,2000ApJ...545..842Q,2001ApJ...561L..77L,2000A&A...362..113F,2002A&A...383..854Y}.
In general models account for the low luminosity of Sgr A* through a very low accretion
rate, through a very low radiative efficiency or through a combination of these.  Estimates of
the accretion rate vary by seven orders of magnitude.  Linear polarization observations can break this
degeneracy.

These new observations complement past work that
has shown a complex picture of the polarization in Sgr A*.  The emission shows
no linear polarization between 1.4 and 112 GHz at limits of 0.1 to 2\% 
\citep{1999ApJ...521..582B,1999ApJ...527..851B,2001ApJ...555L.103B}.  Surprisingly, the
emission is circularly polarized between 1.4 and 43 GHz 
\citep{1999ApJ...523L..29B,1999ApJ...526L..85S,2002ApJ...571..843B}.  The circular polarization is
strong and variable, sometimes showing a highly inverted spectrum.  
Models account for the circular polarization and absence of linear polarization 
through conversion of linear to circular polarization and bandwidth depolarization of
the linear polarization with cold electrons
\citep{2002ApJ...573..485R,2002A&A...388.1106B}.  A model that involves plasma modes
in the vicinity of a black hole has also been developed
\citep{2000AAS...197.8315B}.

The polarization properties above 112 GHz are less certain.  JCMT observations at 150, 220, 375
and 400 GHz (2.0, 1.35, 0.85 and 0.75 mm wavelength, respectively) 
indicate that Sgr A* is polarized at a level of $\sim 10\%$
\citep[][hereafter A00]{2000ApJ...534L.173A}.
However, these observations suffer from the poor angular resolution of the JCMT (22$^{\prime\prime}$ at 220 GHz).  
A substantial amount of polarized
and unpolarized radiation must be subtracted from the flux detected at the Galactic Center
to estimate the polarization of Sgr A*.  
The apparent sharp transition in linear polarization apparent between 112 and 150 GHz
from these observations is very unusual and has important implications for the source physics 
\citep{2000ApJ...545..842Q,2000ApJ...538..L121,2000ApJ...545L.117M}.

We present interferometric observations of the linear polarization in Sgr A* at 230 GHz
(1.3 mm wavelength).   Made with the BIMA array, 
these observations are approximately two orders of magnitude
higher resolution than those of the JCMT.  Thus, we are able to eliminate the effects of
extended polarized and unpolarized emission and isolate the emission of Sgr A* itself.
We present the observations in \S 2, discuss the results in \S 3 and summarize
in \S 4.

\section{Observations and Data Reduction}

Data were obtained on 4 epochs (Table~1) in the B and C configurations of the BIMA array 
\citep{1996PASP..108...93W}.
The measured single sideband system temperature on Sgr A* ranged from 1000 to 2800 K 
for the B-configuration data,
and 1500 to 9000 K for the C-configuration data. 
Both upper (230.7 GHz) and lower sideband (227.3 GHz)
were sampled with a bandwidth of 800 MHz.  
The data were obtained from switched polarization observations 
\citep[see][]{1999ApJ...527..851B}.
The polarization on each antenna is switched between left and right circular
polarization (L and R) using a Walsh switching pattern of period 16.
Using an integration time of 11 second,
all polarization combinations (LL, RR, LR and RL) are sampled every 5 minutes.

The data were phase self-calibrated by using the LL and RR polarizations averaged over 5 minutes.
Phase calibration was applied to all 4 polarizations LL, RR, LR, and RL.
Amplitude calibration of the BIMA array
is determined from measurements of the antenna gains ($\sim 150$ Jy K$^{-1}$)
using planet observations at a number of epochs. 
The antenna gains are stable and
no further amplitude calibration was applied. 

The observed amplitudes are 
sensitive to atmospheric coherence and to antenna pointing errors.  
Both errors affect polarized and unpolarized
emission equally, implying that the polarization fractions are a more reliable
measurement than the total or polarized intensities.
The antenna pointing was checked on the quasar J1733-1304 and Uranus.  
We estimate the magnitude of these errors from measurements of the
flux density for J1733-1304 in February through May 2002. The measurements are
$2.2 \pm 0.2$ Jy at 0.85 mm (JCMT); $2.3 \pm 0.3$ (BIMA) and $3.0 \pm 0.3$ (OVRO),
at 1.3 mm; and, $4.3 \pm 0.3$ (BIMA) and  $3.7 \pm 0.3$ (OVRO) at 3 mm (B. Matthews and
J. Carpenter, private communication).
These are consistent with a coherence loss $< 30\%$ for our 1.3 mm polarization
measurements.


Polarization calibration was applied using the polarization leakage terms
derived from observations in the same mode 
of 3C 273 and 3C 279 on 02 February 2002.  The polarization
leakage for each antenna ranges from $\sim 1$ to 5\% on all but one antenna. The
polarization results for Sgr A* were identical with and without antenna 8, which had
leakage terms $\sim 10\%$.  Comparison of the
polarization calibrations  in 8 epochs over the past 1.5 years 
shows that the leakage terms are stable
and reproducible to  $\sim 1.5\%$ rms 
\citep{2002BIMA..89B}.  This leads to a random systematic error in
the polarization of $\sim 0.5\%$.

We compared the polarized flux density for 3C 279 measured with BIMA at 86 and 230 GHz
and measured with the VLA at 22 and 43 GHz over the past 2 years
\citep{2000VLBA..26T,2002BIMA..89B}.  The fractional polarization shows the same
behavior from 22 to 230 GHz.  The VLA and BIMA position angles differ by $\sim 10$ degrees.  
It is not
known whether this difference is due to source structure or an intrinsic error in the VLA
or BIMA calibration.

The polarization leakage corrections were applied by averaging the switched
polarization data over intervals ranging from 2 to 30 minutes.
We calibrated upper and lower sideband observations separately for each epoch.
The errors in the polarization were consistent with the thermal noise for averaging times
from 5 to 30 minutes.

Combining all the data 
weighted by system temperature and spatially filtered  to include only \uv-data
between 20 and 300 $k\lambda$, we obtain the images shown in Figure~\ref{fig:stokes}. 
With robust
weighting to minimize the noise from sidelobes and thermal noise we obtain a synthesized
beamwidth 3.6 x 0.9 arcsec with a theoretical rms noise level 10 mJy.  We combined
the data in several combinations to look for changes in the polarization with time, frequency
and array configuration.  The results are summarized in Table 2.  Fluxes were found
by fitting point sources at the center of the field.  LSB and USB refer to the lower
and upper sideband data sets including all epochs.  B, C1, C2 and C3 refer to the 
four epochs summarized in Table~1.  C refers to a combination of all three C-configuration
observations.  For the combined result, we find the linear polarization to be
$7.2 \pm 0.6\%$ in position angle $139 \pm 4$ degrees.

\section{Discussion}

\subsection{Dust Polarization}

These observations confirm the existence of high fractional linear polarization in
or near Sgr A* at 230 GHz.   They can be used to eliminate the possibility
of dust contamination of the polarized signal.  A00 estimated 
that their central $22^{\prime\prime}$ beam included 3.55 Jy of (unpolarized)
free-free emission and 0.75 Jy of dust emission at 220 GHz which left $2.2 \pm  0.5$ Jy
for Sgr A*.  A00 assume that
the dust emission is 3\% polarized at a position angle of 100 degrees based on
the larger scale properties of the dust polarization.  If we assume that the
dust and free-free emission are distributed smoothly over this beam, then we expect
a factor of ${{22 \times 22} \over  {3.6 \times 0.9 } }= 150$
decrease in these flux densities for the BIMA result.  Thus, even 100\% polarized
dust emission could produce only $\sim 5$ mJy of the observed 89 mJy of polarized
flux density.

Could the dust be significantly clumped to produce the observed polarization?  
If all of the dust emission were concentrated in our central pixel, then the 
minimum dust polarization fraction is $\sim 12\%$.  However,  this leaves 
only 0.5 Jy flux density for Sgr A* itself, which is very low given the known 
spectrum and history of millimeter flux density observations.  The minimum
flux density observed for Sgr A* at 1.3 mm is $> 1$ Jy 
\citep{2002APJinprepzhao}.  Introducing a 
decorrelation factor $\xi\leq 1$ gives a 
flux density of $0.5 \xi^{-1}$ Jy for Sgr A* and a dust polarization fraction
$12\xi^{-1}\%$.  Our limit on gain errors derived from J1733-1304 places the constraint
$\xi > 0.7$, which doesn't substantially increase the flux density for Sgr A*.
Interferometric observations of dust polarization indicate that
dust polarization rarely exceeds 20\% and only in cases of dense 
star-forming molecular clouds \citep{1998ApJ...502L..75R,2002ApJ...566..925L}. 
An unlikely coincidence would be necessary to place such a rare, highly clumped
and highly polarized cloud at the location of Sgr A*.

We conclude therefore that the effects of dust polarization on our measurement 
are negligible and that the emission is associated with Sgr A* itself.

\subsection{Variations with Wavelength and Time}

While the polarization fraction we measured agrees with that of A00, 
the position angle of $139 \pm 4$ degrees does not.
A00 measured a total polarization of 4.1\% (266 mJy) in position angle $89\pm 3$ degrees in
the central pixel at 220 GHz.  Some of this polarized emission is dust
which A00  assume to be polarized at 3\% (25 mJy) 
in position angle 100 degrees.
We can estimate the actual dust parameters necessary to make our
measurements agree.  Since our measured Sgr A* polarization is almost
orthogonal in $(Q,U)$-space to the central pixel of A00, the
required dust polarization for agreement is extreme:  300 mJy (or 40\%
of the estimated 750 mJy of dust flux density) in position angle
80 degrees.  This high polarization fraction is 
inconsistent with other measured polarization fractions as discussed above.  
Thus, we cannot easily reconcile the A00 220 GHz result with ours
through the dust properties.

We can estimate the rotation measure from the position angle data as a function of
frequency (Figure~\ref{fig:rmplot}).
The lack of bandwidth depolarization in the
BIMA results places a strong upper limit of $\sim 10^8 \rdm$ on the RM.  
The LSB and USB results for BIMA are nearly identical, placing
an upper limit of $2 \times 10^6 \rdm$.  The best fit to these data points
is $-3 \times 10^5 \rdm$ but the error is $>10^6 \rdm$.  
Although we can reconcile all the BIMA and JCMT 
position angles with RM $\sim -3 \times 10^6 \rdm$ and three phase wraps between 150 and 400
GHz, a RM this large would lead
to substantial depolarization in the broad (40 GHz) JCMT bandpasses.  We would expect to find
a much higher polarization fraction in the BIMA results which have much narrower
bandpasses.  In fact, based on the nearly equal polarization
fractions in the BIMA and JCMT results, we can conclude again that the polarization must be $< 10^6
\rdm$.

However,
if we exclude the A00 220 GHz result, we find that all of the 
remaining polarization detections can be explained as the result of a single RM
without leading to substantial depolarization.  We measure
${\rm RM}=-4.3 \pm 0.1 \times 10^5 {\rdm}$ and a zero-wavelength position angle of
$181 \pm 2$ degrees.  Including the A00 220 GHz result gives 
$-4.3 \times 1.6 \times 10^5 \rdm$.  This error could be construed as including
the effects of systematic measurement error and variability.
If we exclude both the 150 and 220 GHz A00 result,  
we find ${\rm RM}=-4.4 \pm 0.4 \times 10^5 {\rdm}$.  While the exclusion of the low
frequency JCMT points in the RM fit appears {\it ad hoc}, note that these are the least
reliable of the JCMT results due to the much larger beam size.  Nevertheless, this
RM measurement must be considered tentative.

We cannot eliminate variability as an explanation for the differences 
in position angle that are seen.  The A00 results were obtained in three separate epochs
spread over 5 months nearly three years earlier than the BIMA results.  
However, we detect no change in the position angle of the linear polarization with 
time in the BIMA data (Figure~\ref{fig:results}).  
The reduced $\chi^2$ is 0.5 for the four epochs under the hypothesis of a 
constant position angle.  Further, the A00 measurements at 375 and 400 GHz were separated
by 5 months but show no apparent change.  On the other hand, the fractional linear polarization in
the BIMA data is slightly variable.
The final epoch C3 is significantly different from the mean.  
The reduced $\chi^2$ is 2.0 for
the four epochs under the hypothesis of a constant polarization.  

We summarize detections and upper limits for the linear polarization fraction
in Figure~\ref{fig:pfrac}.  There is a sharp transition in the polarization 
fraction between 100 and 200 GHz.  Our previous BIMA measurement that led to an 
upper limit at 112 GHz cannot be explained by bandwidth depolarization.
Bandwidth depolarization occurs at 112 GHz in 800 MHz of bandwidth when
${\rm RM} > 1.0\times 10^7 \rdm$.  

On the other hand, angular depolarization, 
which requires that the RM change the position angle by $>180$ degrees across the
face of the source, is marginally adequate.
At 112 GHz, an RM a few times $10^5 \rdm$ is just sufficient.
Thus, a fully turbulent accretion region with
an outer scale comparable to the source size ($< 1 {\rm AU}$, see discussion below)
could depolarize the source.   However, measured RM flucutuations in the Galactic Center and
Cygnus regions have a maximum of $<10^3 \rdm$ on degree scales and are $<<1 \rdm$
when extrapolated to scales below 1 arcsec \citep{1990ApJ...363..515L,1997ApJ...475L.119Y}.
Nevertheless, turbulence in the accretion environment may be substantially different
from that in the interstellar medium.

An alternative 
explanation is that Sgr A* consists of an unpolarized source and a 
polarized source with a steeply inverted spectrum that dominates the spectrum 
above 230 GHz.  
A two component model is natural for both inflow and outflow models since they 
must account for the sharply increasing spectrum at millimeter wavelengths
\citep{1997ApJ...490L..77S,1998ApJ...499..731F}.  All
models (see next section) that fit the centimeter to submillimeter spectrum include at least two
distinct populations of radiating particles.  If the spectrum rises as $\nu^{2.5}$, 
then the source would depolarize
by a factor of 0.16 between 230 and 112 GHz, which is sufficient to account for 
the absence of detection at 112 GHz.  Spectral indices of $<1$ and $>4$ are
excluded by the results.

\subsection{Circular Polarization}

Circular polarization is not detected in any of these data sets.  For the cumulative
data set the circular polarization is less than the thermal rms noise of $\sim 1\%$.  Only in the B array
data is there a hint of a detection.  Here the circular polarization is two times the
noise.

\subsection{Origin and Propagation of the Linear Polarization}

The maximum observed RM in the Galactic
Center is on the order of a few times $10^3 \rdm$ \citep{1997ApJ...475L.119Y}.
However, this occurs in a nonthermal filament approximately 0.5 degrees away from Sgr A* and may not probe the 
densest ionized regions towards the Galactic Center.
A hyperstrong interstellar scattering screen significantly broadens the image of Sgr A* at 
all wavelengths \citep{1998ApJ...508L..61L,1998ApJ...505..715L,2001ApJ...558..127B}.
The physical parameters of this scattering screen are not fully constrained.  However,
we can estimate an upper limit to the RM of the screen.  The maximum RM is  $\sim 10^4 \rdm$ 
for models of the ionized surfaces of molecular clouds \citep{1994ApJ...434L..63Y,
1998ApJ...505..715L,1999ApJ...521..582B}.  However, the RM may arise in a region different
from the scattering region.  Sgr A* is embedded in or behind Sgr A West as well as a dense,
ionized halo that extends over $\sim 5^{\prime}$ 
\citep{1989ApJ...342..769P,1999cpg..conf..422A}.  This halo has density of 
$10^2 - 10^3 {\rm\ cm^{-3}}$ and a scale of 10 pc.  For magnetic field strengths 
of a milliGauss, the maximum RM is $\sim 10^7 \rdm$.  Similarly, Sgr A West 
has a density $10^2 {\rm\ cm^{-3}}$, a scale of 1 pc and milliGauss field strength,
giving a maximum RM $\sim 10^5 \rdm$.
Thus, while the interstellar scattering screen does not contribute significantly to the
measured RM, material which closer to Sgr A* but not directly associated with accretion and
outflow could.

Millimeter VLBI observations require a size for Sgr A* of 72 gravitational radii or less at 
43 GHz and sizes decreasing as $\sim \nu^{-1}$ at higher frequencies
\citep{1998ApJ...496L..97B,1998ApJ...508L..61L,1998A&A...335L.106K,2001AJ....121.2610D}.
Theoretical temperature profiles indicate
that the emission comes from a volume within a few gravitational radii of Sgr A*
\citep{2001ARA&A..39..309M}.
In ADAF or CDAF models,
the 230 GHz source is embedded within a quasi-spherical accretion region.  Imaging
by Chandra at X-ray wavelengths indicate that Sgr A* is extended on a scale 
consistent with the radius of this larger region, suggesting a concentration
of hot thermal plasma \citep{2001AAS...199.8503B}.  

The emission must propagate through the accretion region without substantial depolarization.
Using the measured RM $\sim -4\times 10^5 \rdm$, we can estimate the accretion rate.
(Note that the following conclusions also hold for the upper limit RM of 
$2\times 10^6 \rdm$  determined from the BIMA data).
The implied accretion rates differ between models because of
the different temperature versus radius relations.
We can calculate the Bondi-Hoyle accretion rate assuming equipartition and a uniform
magnetic field equation (5) from \citet{1999ApJ...521..582B}.
To produce the measured RM, the accretion rate must be $10^{-7} \msuny$.
However, this 
accretion rate is too low to produce the observed spectrum by two orders of
magnitude for both the ADAF and Bondi-Hoyle models.  
An accretion rate of $10^{-5} \msuny$ for these models
can model the spectrum but requires an RM $\sim 10^{10} \rdm$.
On the other hand, CDAF models can  model the spectrum with 
$\dot{M} \sim 10^{-8} \msuny$, which produces ${\rm RM} \sim 10^6 {\rdm}$
\citep{2000ApJ...545..842Q}.  Jet models require an accretion rate of $10^{-8} \msuny$,
which is consistent with a low RM \citep{2000A&A...362..113F}.
\cite{2000ApJ...545L.117M} predict $10^{-11} \msuny$, which significantly 
underpredicts the RM.  This model requires that the RM originates 
external to Sgr A*.

Two effects may mitigate the conclusions discussed above.  These are magnetic field
reversals and deviations from equipartition between particle and magnetic field energy.
If the magnetic field reverses frequently, then the rms field in the region is 
a better indicator of the characteristic field for a RM calculation
\citep{2002ApJ...573..485R}.  A model predicting a high RM for a uniform field
may produce an RM lower by orders of magnitude if there are many field reversals.
However, the number of field reversals necessary to account for the difference between
the ADAF or Bondi-Hoyle models and our measured RM is on the order of $10^8$, which
is extreme.
Reducing the magnetic field strength substantially below equipartition
values will reduce the RM by the same factor.  However, these reductions will be offset
by the necessary increase in particle density to model the same spectrum.

In jet models, the emission originates in a region of similar scale that can be located
an arbitrary distance from the black hole.  Jet emission is likely to originate in
a compact shocked component near the base of the jet
\citep{2002A&A...383..854Y}.  The compressed magnetic field
will be perpendicular to the axis of the jet, implying that the jet is oriented
North-South.  Increasing magnetic field order towards the base of the jet can
account for the sharply increasing polarization fraction with frequency.
High frequency
VLBI observations of quasars and BL Lac objects indicate that electric vector position
angles are typically but not always aligned with the jet axis \citep{2001ApJ...562..208L}.
The inclination angle of a jet in Sgr A* is not well-constrained.  However,
an angle close to the line of sight is possible given that highly inclined sources often
produce the highest linear polarization 
in compact radio sources.  This raises the possibility that
the radio/millimeter emission is relativistically beamed, which may also account for
the absence of a visible jet or ejected components despite strong variability 
\citep[e.g.,][]{1998ApJ...496L..97B}.

The jet model does not require propagation through the accretion medium.  However,
the high value for the RM suggests that it may pass through some of the accretion or outflow
medium \citep{2002A&A...388.1106B}.  In AGN cores, the maximum RMs measured are on the order of 2000 $\rdm$
\citep{2000ApJ...533...95T}.  This argues that we are probing a region in Sgr A* quite different 
from that in other AGN.

The RM we measure is not inconsistent with a lower limit 
for the RM from spectropolarimetry at 4.8 and 8.4 GHz \citep{1999ApJ...521..582B}.
No linear polarization was detected for RMs as high as $10^7 \rdm$ in
this experiment.  Different source physics may account for the absence
of polarization at the longer wavelengths.  This will result if the lower
frequency source is intrinsically depolarized due to greater internal field disorder,
for example.  Alternatively, 
foreground beam depolarization across the much larger lower frequency source 
may account for the absence of linear polarization.

\subsection{Predictions for Future Observations}

An important expectation of emission from a jet is variable linear polarization
\citep{1985ApJ...298..114M}.  Variable polarization may
also be a measure of the stability of accretion in a disk model.
For example, \citet{2002ApJ...573L..23L} have suggested that precession of the
disk on a timescale of 100 days is a consequence of a very low value for the
spin of the black hole.  Such precession could lead to a change in the 
position angle of the polarization if the disk orientation is linked to the
magnetic field structure.
We have not seen convincing evidence for this position angle
variation yet, except perhaps in the
discrepancy between A00 and our result.  
Significant variability in the total intensity at centimeter, millimeter, 
submillimeter and X-ray wavelengths on timescales of hours to years has been reported 
\citep{1993ApJ...417..560W,2001AAS...19910309Y,2001Natur.413...45B,2001ApJ...547L..29Z,2002ApJ...571..843B}.
Correlating polarization variability with these events
could be an important clue to the origin of the polarization.
However, the stable polarization that we have measured over 60 days
and that A00 measured over 5 months between 375 and 400 GHz
suggests that such connections are possibly subtle or require
intensive monitoring.  If changes in the polarization are associated with X-ray flares,
then they may have a timescale of hours.   

The detection of a RM and a power law index for the 
polarization fraction can be tested with observations at other
wavelengths.
BIMA observations can be spread more widely over frequency space.  The
1.3 mm receivers are sensitive from 210 to 270 GHz.  With properly
calibrated quarter wave plates, BIMA could measure the expected
difference in position angle of $\sim 20$ degrees over this range.
BIMA can also measure differences in the power law index for the
polarization fraction over this same range.  Interferometric 150 GHz
observations should see a substantially decreased polarization fraction
but a significantly different polarization angle than observed 
at 230 GHz.  Higher frequency observations may see the polarization
fraction continue to increase.

Variability in the RM can be used as a diagnostic of the accretion 
or outflow medium.  Sgr A* shows variability on all timescales
from hours to years at radio wavelengths 
\citep{2001ApJ...547L..29Z,2002ApJ...571..843B}.  This variability
may be indicative of changes in the accretion or outflow rates
or conditions, which may be reflected in the RM.  
Refractive scintillation timescales are on the order
of 1 month at 230 GHz for a velocity of 100 ${\rm\ km s^{-1}}$.

The detection of linear polarization raises the possibility
of polarimetric VLBI.  Images
made at 230 GHz or higher frequencies will be only weakly
affected by scattering and may show the effects
of general relativity 
\citep{2000ApJ...528L..13F,2000AAS...197.8315B,2001ApJ...555L..83B}.
This may be the best probe of the close environment of a black hole
available.


\section{Conclusions}

We have confirmed the existence of strong linear polarization at millimeter
wavelengths in Sgr A* with high resolution observations by the BIMA
array.  These eliminate the possibility of confusion from polarized
dust emission.  The polarized emission probably comes from within 10
gravitational radii of the black hole.  This is one of the closest probes
of the environment of a black hole that exists.

Propagation of the polarized radiation through the accretion and outflow
media provides important limits on the physical characteristics in 
these regions.  The measured RM appears to favor a very low accretion
rate which excludes standard ADAF models and Bondi-Hoyle models with
very high accretion rates for Sgr A*.

The existence of linear polarization at these high frequencies
opens up several new methods for investigation of this source.  Variations
in polarization angle and fraction as well as rotation measure
may provide critical constraints on models.  Very long baseline
interferometric imaging of linearly
polarized emission in the vicinity of the black hole may reveal
a wealth of effects, including those of general relativity.

\acknowledgements 
The BIMA array is operated by the Berkeley-Illinois-Maryland Association under
funding from the National Science Foundation. 


\plotone{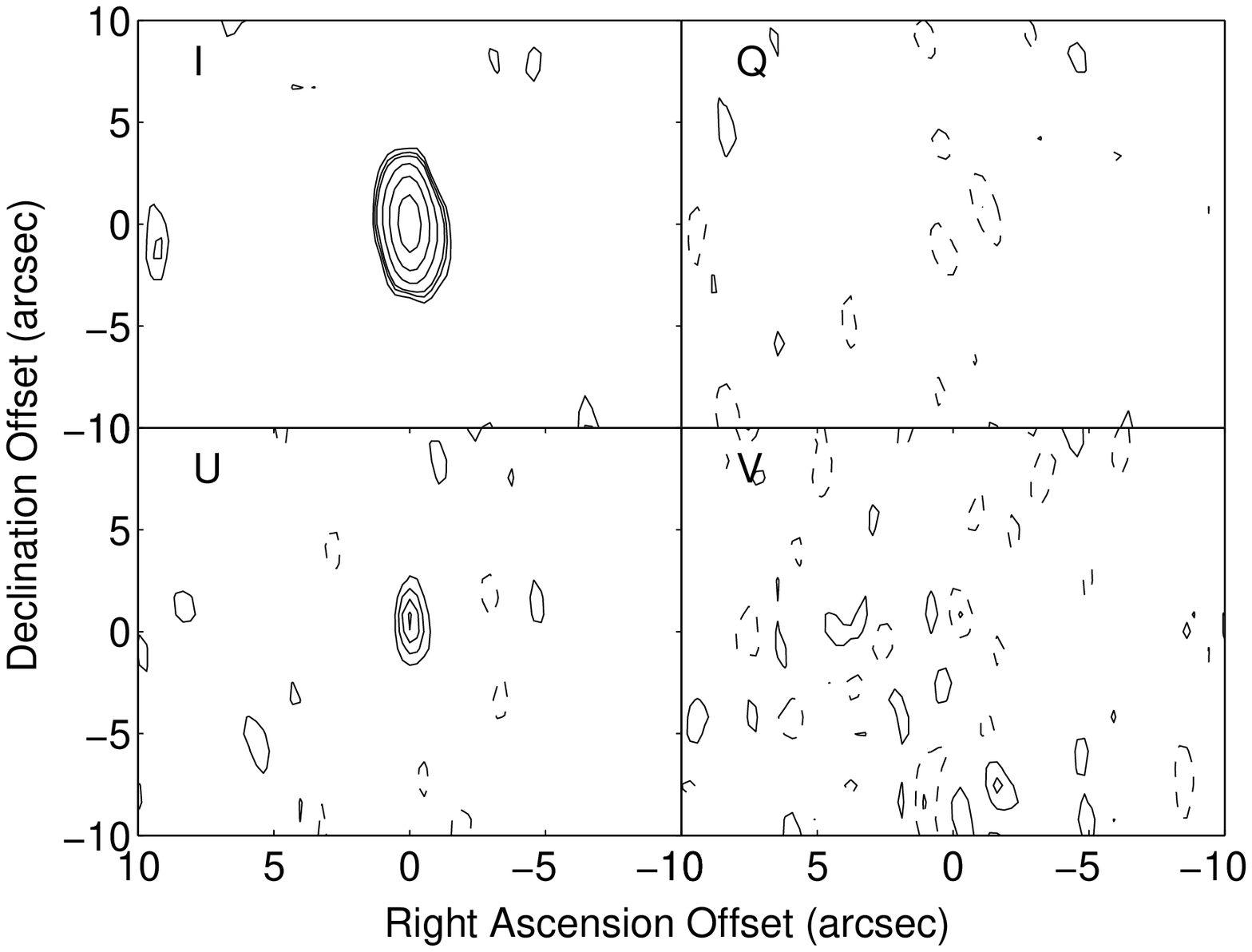}
\figcaption[f1.eps]{Image of Sgr A* in Stokes I, Q, U and V from the combined
data set of all observations.  The synthesized beam is
3.6 by 0.9 arcsec.  Contours are 2, 4, 6 and 8 times the rms noise of 11 mJy beam$^{-1}$ in
the Q, U and V images.  In the polarization images, 
positive and negative contours are shown as dashed and solid lines, 
respectively. Contours are shown as solid lines at 
4, 6, 8, 16, 32, and 64 times 11 mJy beam$^{-1}$ in the
I image.\label{fig:stokes}}

\plotone{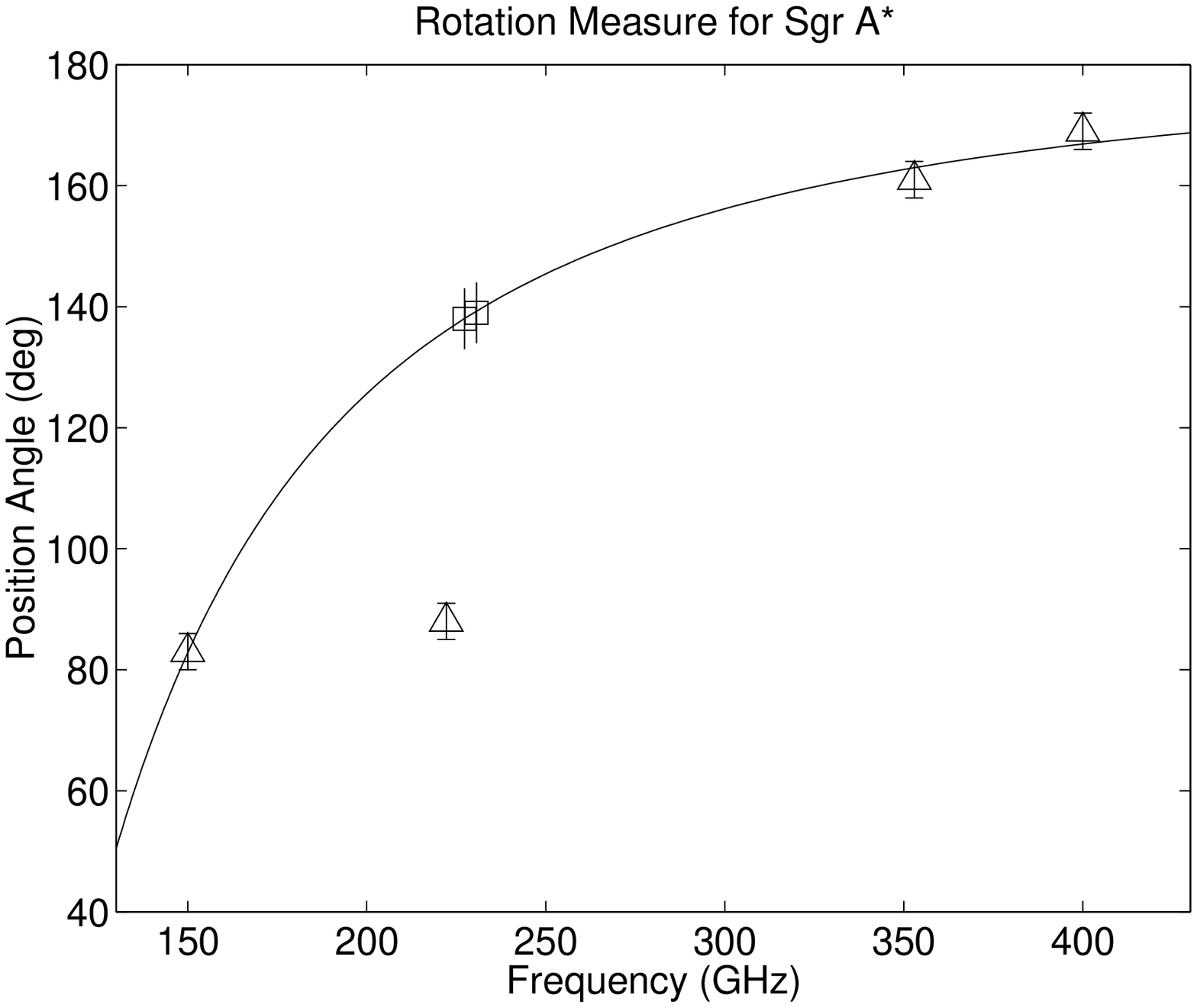}
\figcaption[f2.eps]{
Position angle as a function of frequency.  Triangles are
the A00 data.  Squares are the BIMA data.  The solid line is a fit for the RM
excluding the A00 230 GHz result.  The best fit is $-4.3 \pm 0.1 \times 10^5 \rdm$ with
a zero-wavelength position angle of $181 \pm 2$ degrees.
\label{fig:rmplot}}

\plotone{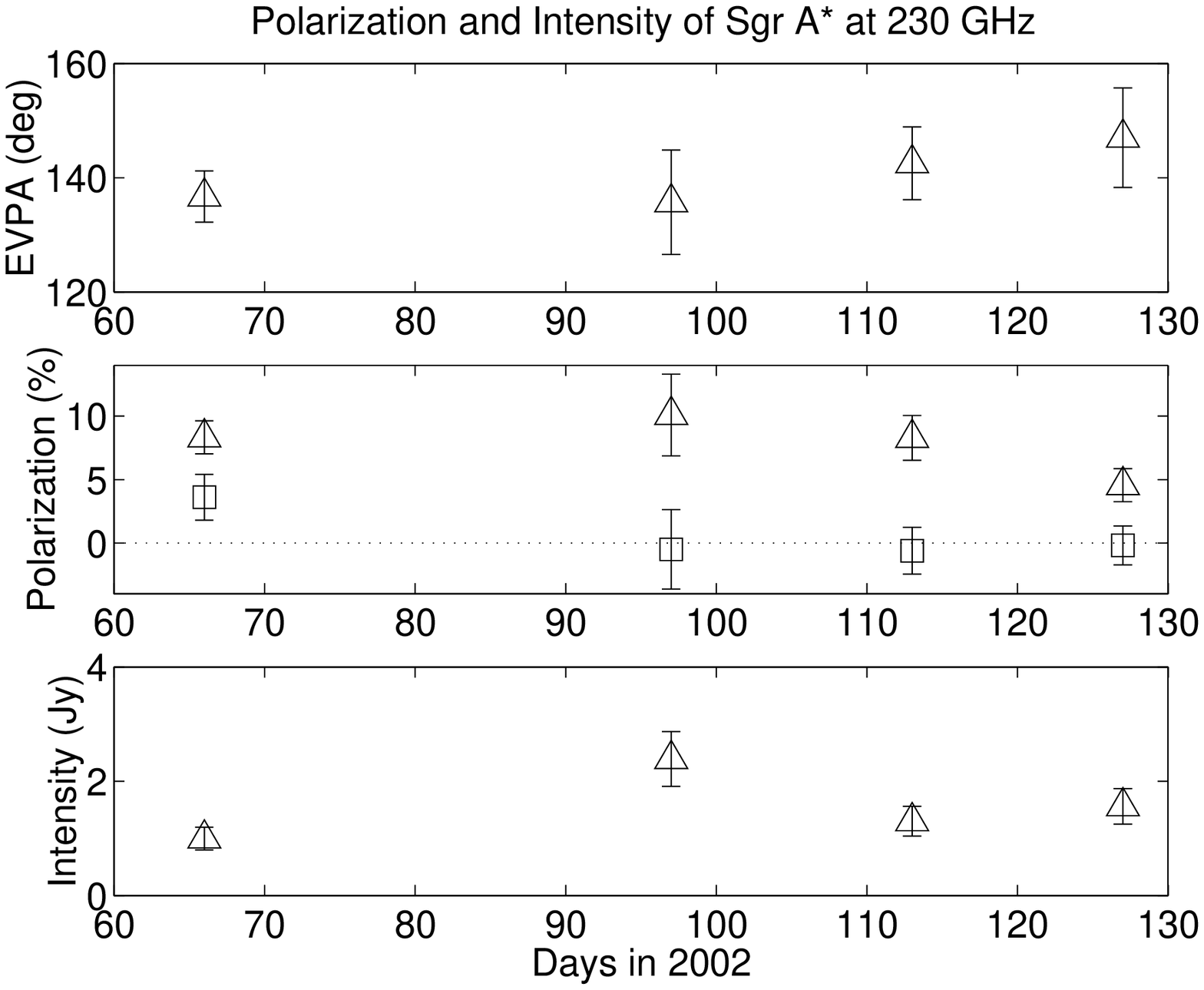}
\figcaption[f3.eps]{
Lightcurves of the linear and circular polarization of Sgr A* as
measured with BIMA at 230 GHz.  In the upper panel, the electric vector position angle
is plotted.  In the middle panel,  the fractional linear (triangles) and circular
(squares) polarization is plotted.  In the lower panel, the total intensity is
plotted.
\label{fig:results}}

\plotone{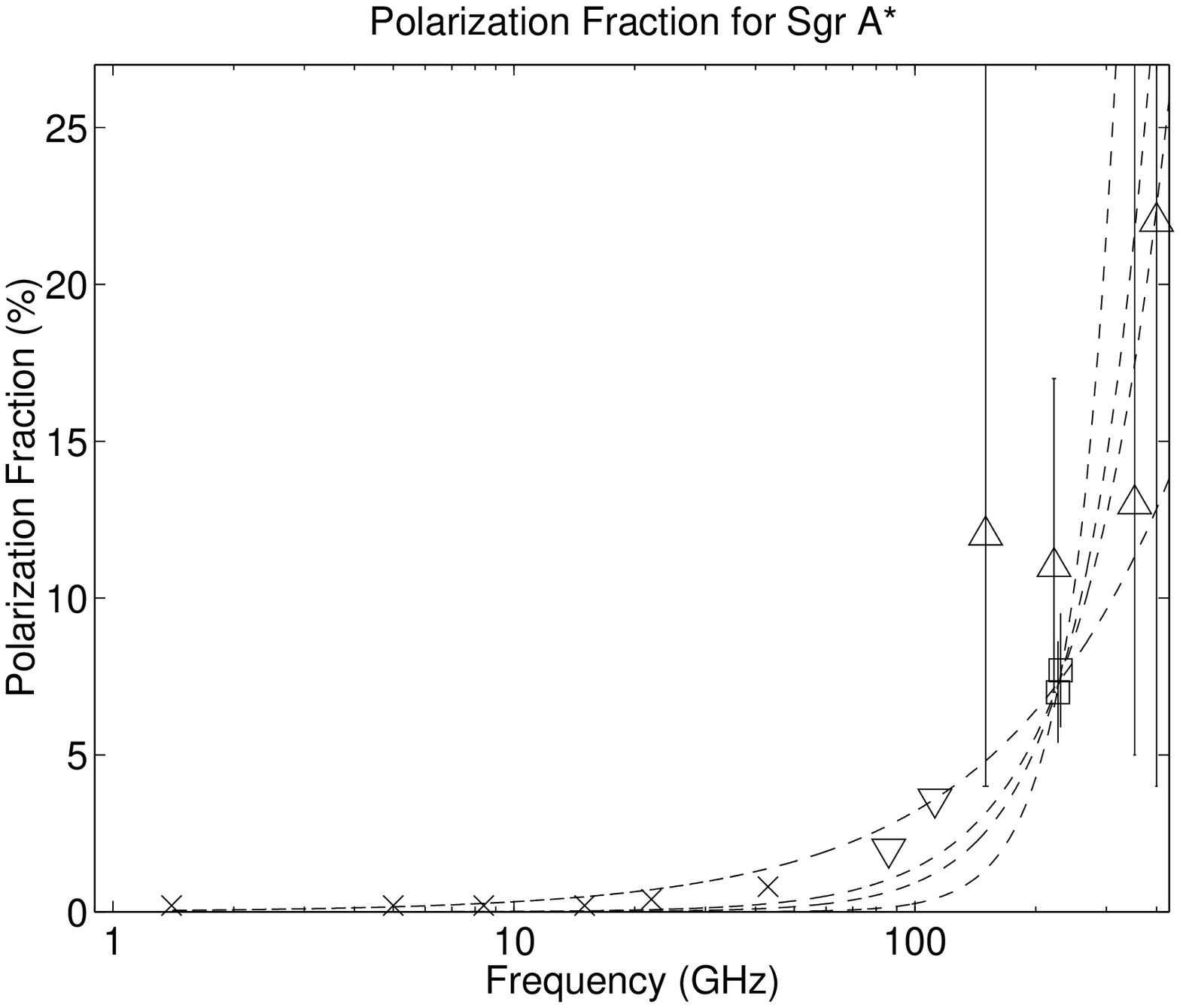}
\figcaption[f4.eps]{
Polarization fraction as a function of frequency.  Triangles
are the A00 data.  Squares are the new BIMA data.  The inverted 
triangles are the 3 mm BIMA upper limits.  Crosses are VLA upper limits.  All error bars
and upper limits are plotted at 2$\sigma$ levels.  The dashed lines are power law models
for the polarization fraction with indices of 1.0, 2.0, 2.5 and 4.
\label{fig:pfrac}}

\begin{deluxetable}{rcrrrr}
\tablecaption{Polarimetric Observations of Sgr A*}
\tablehead{\colhead{Date} & \colhead{Array} & \colhead{$b_{maj}$} & \colhead{$b_{min}$} & \colhead{p.a.} & \colhead{$\tau_{230}^{zenith}$}  
\\
                &       & \colhead{($^{\prime\prime}$)}  & \colhead{($^{\prime\prime}$)}  & \colhead{(deg)} & 
\colhead{(nepers)}                }
\startdata
02 March 2002 & B & 2.6 & 0.7 & 1.7 & 0.29 \\
08 April 2002 & C & 6.5 & 1.7 & 9.4 & 0.78 \\
24 April 2002 & C & 8.7 & 1.7 & 3.2 & 0.65 \\
08 May   2002 & C & 6.6 & 1.8 & 8.7 & 0.36 \\
\enddata
\end{deluxetable}

\begin{deluxetable}{lrrrrrr}
\tablecaption{Polarized and Total Flux Density of Sgr A* at 230 GHz}
\tablehead{\colhead{Data Set} & \colhead{$I$} & \colhead{$Q$} & \colhead{$U$} & \colhead{$V$} & \colhead{$p$} & \colhead{$\chi$} \\
                    & \colhead{(mJy)}& \colhead{(mJy)} &\colhead{(mJy)} &\colhead{(mJy)} & \colhead{(\%)} & \colhead{(deg)} }
\startdata
ALL  & $ 1235 \pm   24 $& $   12 \pm   11 $& $  -88 \pm   11 $& $   13 \pm   14 
$& $  7.2 \pm  0.6 $& $  139 \pm    4 $\\
\cline{1-7}
\cline{1-7}
LSB  & $ 1344 \pm   26 $& $   10 \pm   15 $& $  -93 \pm   16 $& $   19 \pm   19 
$& $  7.0 \pm  0.8 $& $  138 \pm    5 $\\
USB  & $ 1165 \pm   27 $& $   13 \pm   15 $& $  -89 \pm   16 $& $   -2 \pm   18 
$& $  7.7 \pm  0.9 $& $  139 \pm    5 $\\
\cline{1-7}
\cline{1-7}
B    & $  999 \pm   23 $& $    5 \pm   13 $& $  -83 \pm   13 $& $   36 \pm   18 
$& $  8.3 \pm  0.9 $& $  137 \pm    4 $\\
C    & $ 1483 \pm   38 $& $   27 \pm   16 $& $  -90 \pm   17 $& $    0 \pm   18 
$& $  6.3 \pm  0.8 $& $  143 \pm    5 $\\
C1   & $ 2389 \pm   89 $& $    6 \pm   77 $& $ -241 \pm   77 $& $  -12 \pm   75 
$& $ 10.1 \pm  2.3 $& $  136 \pm    9 $\\
C2   & $ 1301 \pm   59 $& $   28 \pm   24 $& $ -104 \pm   23 $& $   -8 \pm   24 
$& $  8.3 \pm  1.3 $& $  143 \pm    6 $\\
C3   & $ 1561 \pm   45 $& $   29 \pm   22 $& $  -65 \pm   20 $& $   -3 \pm   24 
$& $  4.6 \pm  0.9 $& $  147 \pm    9 $\\
\enddata
\end{deluxetable}
\end{document}